\documentclass[a4paper]{llncs}

\usepackage{amsmath}
\usepackage{amssymb}
\usepackage{graphicx}
\usepackage{algorithm2e}
\usepackage{breqn}
\usepackage{multirow}
\usepackage{capt-of}
\usepackage[T1]{fontenc}
\usepackage{letltxmacro}

\graphicspath{{Figures/}}
   \DeclareGraphicsExtensions{.pdf,.jpg,.png}

\begin{document}

\title{Mining Software Components from Object-Oriented APIs\vspace{-1.0em}}

\author{Anas Shatnawi\inst{1} \and Abdelhak Seriai\inst{1}
 \and Houari Sahraoui\inst{2} \and Zakarea Al-Shara\inst{1}}

\institute{UMR CNRS 5506, LIRMM, University of Montpellier II, Montpellier, France\\
\email{{shatnawi, seriai, alshara}@lirmm.fr}
\and
DIRO, University of Montreal, Montreal, Canada\\
\email{sahraoui@iro.umontreal.ca\vspace{-1.8em}}
}

\maketitle
\begin{abstract}
\vspace{-0.5em}
Object-oriented Application Programing Interfaces (APIs) support software reuse by providing pre-implemented functionalities. Due to the huge number of included classes, reusing and understanding large APIs is a complex task. Otherwise, software components are admitted to be more reusable and understandable entities than object-oriented ones. Thus, in this paper, we propose an approach for reengineering object- oriented APIs into component-based ones. We mine components as a group of classes based on the frequency they are used together and their ability to form a quality-centric component. To validate our approach, we experimented on 100 \textit{Java} applications that used \textit{Android} APIs.
\vspace{-1.4em}
\end{abstract}

\vspace{-1.20em}
\section{Introduction}
\vspace{-0.80em}
Nowadays, the development of large and complex software applications is based on reusing pre-existing functionalities instead of developing them from scratch \cite{Frakes:2006:IEEETrans,Zibran:2011:WCRE}. Application Programming Interfaces (APIs) are recognized as the most commonly used repositories supporting software reuse \cite{Frakes:2006:IEEETrans}. APIs provide a pre-implemented, tested and high quality set of functionalities \cite{Zibran:2011:WCRE,Monperrus:2012:ESE}. Consequently, they increase software quality and reduce the effort spent on coding, testing and maintenance activities \cite{Zibran:2011:WCRE}.

In the case of object-oriented APIs, e.g., \textit{Standard Template Libraries} in \textit{C++} or \textit{Java SDK}, the functionalities are encapsulated as object-oriented classes. It is well known that reusing and understanding large APIs such as \textit{Java SDK}, which contains more than 7.000 classes, is not an easy task \cite{Homan:2006:APSEC,Uddin:2012:TAA}. Consequently, several approaches have been proposed, such as \cite{Wang:2013:MSH,Montandon:2013:WCRE,Monperrus:2010:ECOOP}, in order to facilitate the understandability and the reusability of APIs by discovering frequent usage patterns of APIs. This is based on the API usage history of software applications (i.e. API clients). Despite the value of frequent usage patterns, these are not sufficient to provide a high degree of API reusability and understandability. These are used as guides for reusing API classes and are not themselves reusable entities \cite{Maalej:2013:IEEETrans}.

Otherwise, software components are admitted to be more reusable and understandable entities than Object-Oriented (OO) ones \cite{szyperski:2002:component}. This is because components are considered coarse-grained software entities, while OO classes are considered fine-grained ones. In addition, components define their required and provided interfaces. This means that the component dependencies are more understandable compared to the dependencies among objects.
Consequently many approaches have been proposed to identify components from OO software applications such as \cite{Kebir:2012:WICSA,allier:2011:WICSA}. Nevertheless, no approach has been proposed to identify components from object-oriented APIs.
Thus, in this paper, we propose an approach to mine components from object-oriented APIs. This does not only improve the reusability of APIs themselves, but also supporting component-based reuse techniques by providing component based APIs. The approach exploits specificities of API entities. We statically analyze the source code of both APIs and their software clients to identify groups of API classes that are able to form OO components. This is based on two criteria. The first one is the probability of classes to be reused together by API clients.
The second one is related to the structural and behavioral dependencies among classes and thus their ability to form a quality-centric component. In order to validate the proposed approach, we experimented on a set of 100 \textit{Java} applications that use three \textit{Android} APIs. The evaluation shows that structuring object-oriented APIs as component-based ones improves the reusability and the understandability of these APIs.

The rest of this paper is organized as follows. The subsequent section, Section 2 puts in context the problem of component identification from APIs. It presents the goal of the proposed approach, the background needed to understand our proposal and the problem analysis. Section 3 presents the foundation of our approach. Then, in Section 4 we present the identification of component interface classes. Section 5 presents how APIs are organized as component-based libraries. Experimentation and results of our approach are discussed through three APIs case studies in section 6. Next, the related work is discussed in Section 7. Finally, concluding remarks and future directions are presented in section 8.
\vspace{-1.40em}
\section{Putting Problem in Context}
\vspace{-0.80em}
\subsection{The Goal: Object to Component}
\vspace{-0.80em}
Our goal is to reengineer object-oriented APIs into component based ones. Based on \cite{szyperski:2002:component,luer:2002:composition,heineman:2001:component}, we consider a component as,  \textquotedblleft a software element that (a) can be composed without modification, (b) can be distributed in an autonomous way, (c) encapsulates the implementation of one or many closed functionalities, and (d) adheres to a component model\textquotedblright. According to this definition, we  derive three quality characteristics that should be satisfied by a component; \textit{Composability}, \textit{Autonomy} and \textit{Specificity}.
\textit{Composability} of a component refers to its ability to be composed through its interfaces without any modification. \textit{Autonomy} refers to that a component can be reused in an autonomous way because it encapsulates the strongly dependent functionalities. \textit{Specificity} refers to that a component implements a limited number of closed functionalities, which makes it a coarse-grained entity. Based on that, we consider as OO components those implemented as a group of OO classes. 

\begin{figure}[h]
\vspace{-0.7em}
  \begin{center}
	\includegraphics[scale=0.50]{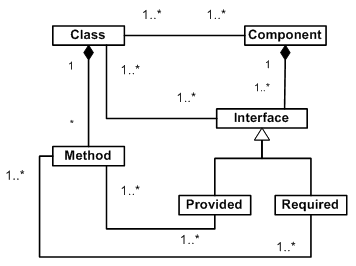}
    \caption{Mapping object to component}
        \label{fig:objectcomponent}
  \end{center}
  \vspace{-2.5em}
\end{figure}

In the context of our approach, the identification\footnote{Component identification is the first step of the migration process of object-to-component} of a component means identifying OO classes that can be considered as the implementation of this component. Thus we consider that a component can be identified from a cluster of classes that may belong to different packages.  Classes that have direct links (e.g. method call, attribute access) with classes implementing other components compose the interfaces of the component. Provided Interfaces of a component are defined as a group of methods implemented by classes composing these interfaces. Required interfaces of a component are defined as a group of methods invoked by the component and provided by other components.  Figure \ref{fig:objectcomponent} shows our object to component mapping model.

\subsection {Background}
\vspace{-0.4em}
\subsubsection{Identifying Components in Software Applications : Synthesis of Previous Work}
\label{ROMANTIC}
\begin{figure}[h]
  \vspace{-2.0em}
  \begin{center}
    \includegraphics[scale=0.30]{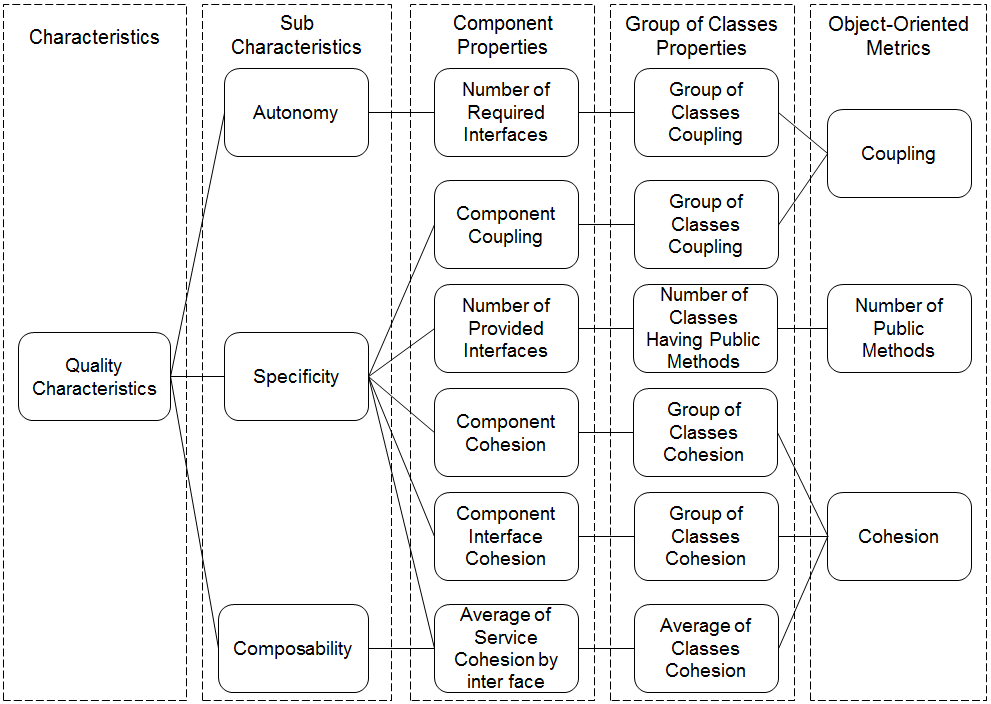}
    \caption{From component characteristics to object-oriented metrics}
        \label{fig:componentquality}
  \end{center}
  \vspace{-3.1em}
\end{figure}

We have proposed in our previous works related to ROMANTIC\footnote{ROMANTIC: Re-engineering of Object-oriented systeMs by Architecture extractioN and migraTIon to Component based ones} approach \cite{Kebir:2012:WICSA,Chardigny:2008:WICSA} a set of metrics to measure the ability of a group of classes in a software application to form a component. These metrics are defined based on the main characteristics of a component (i.e. \textit{Composability}, \textit{Autonomy} and \textit{Specificity}). Similar to the software quality model ISO 9126 \cite{ISO9126}, we proposed to refine the characteristics of the component into sub-characteristics. Next, the sub-characteristics are refined into the properties of the component (e.g. number of required interfaces). Then, these properties are mapped to the properties of the group of classes from which the component is identified (e.g. group of classes coupling). Lastly, these properties are refined into OO metrics (e.g. coupling metric). This quality refinement model is shown in Figure \ref{fig:componentquality}. According to this model, a quality function has been proposed to measure the component quality. This quality function is used as a similarity metric for a hierarchal clustering algorithm \cite{Kebir:2012:WICSA,Chardigny:2008:WICSA} as well as in search-based algorithms \cite{Chardigny:2008:ECSA} to partition the OO classes into groups; where each group represents a component.

\vspace{-1.5em}
\subsubsection{Frequent Usage Patterns}
In the domain of data mining, a Frequent Usage Pattern (FUP) is defined as a set of items, subsequences or substructures that are frequently used together by customers \cite{han2006data}. It provides information that helps decision makers (e.g. customer shopping behavior) by mining associations and correlations among a set of items in a huge data set. An example of FUP mining is a market basket analysis. In this example, the customer buying habits are analyzed to identify items that are frequently bought together in the customer shopping baskets, for instance, milk and bread form a FUP when they bought frequently together. The identification of FUP is based on \textit{Support} quality metric that is used to measure the interestingness of a set of items. \textit{Support} refers to the probability of finding a set of items in the transactions. For example, the value of 0.30 \textit{Support}, means that 30\% of all the transactions contain the target item set. The following equation refers to \textit{Support}:
\vspace{-0.5em}
\begin{equation}
\label{EqSupport}
S(E1, E2) = P(E1 \cup E2)
\end{equation}
Where \textit{E1}, \textit{E2} are sets of items; \textit{S} refers to \textit{Support}; \textit{P} refers to the probability.

\vspace{-1.1em}
\subsection{Component and Frequent Usage Pattern}
  \vspace{-0.5em}
FUPs are observations made based on the analysis of previous uses of APIs. They aim to help users of APIs by identifying recurring patterns, composed of classes frequently used together.

FUPs and components serve the reuse needs in two different ways. Components are entities that can be directly reused and integrated into software applications, while FUPs are guides for reuse and not entities for reuse.
In addition, components and FUPs are structurally different. Related to \textit{Specificity} characteristic, classes composing a component serve a coherent body of services, while classes composing a FUP may be related to different services. Concerning \textit{Autonomy} characteristic, dependencies of component's classes are mostly internal, which forms an autonomous entity. FUP's classes can be very dependent on other classes that are not directly used by clients of APIs. Concerning \textit{Composability} characteristic, a component is structured and reused via interfaces, while FUPs are not directly reusable entities.

\vspace{-1.0em}
\section{The Proposed Solution Foundations}
\begin{figure}[h]
  \vspace{-0.9em}
  \begin{center}
    \includegraphics[scale=0.35]{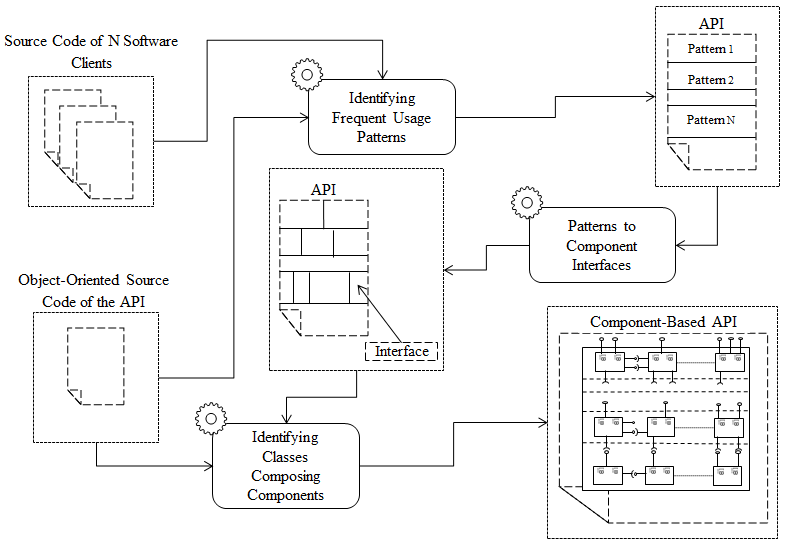}
    \caption{The process of mining components from an object-oriented API}
        \label{fig:miningprocess}
  \end{center}
\end{figure}

  \vspace{-1.0em}
Based on the observations made in the previous sections, we consider that:
  \vspace{-0.5em}
\begin{itemize}

\item	In object-oriented APIs, a component is identified as a group of classes.

\item	To reengineer the entire object-oriented API into component-based one, each class of the API is mapped to be part of at least one component. Each class is mapped either as a class of the component interfaces or as a part of the internal classes of the component.

\item	Classes directly accessed by the software clients represent the end-users' services. These classes compose FUPs. These ones are the candidate to form the provided interface of the components mined from the API.

\item	As a FUP can be composed of classes providing multiple services, its classes can be mapped to be a part of different component interfaces.

\item	A class of an API can be a part of several FUPs and can participate to implement multiple services. Consequently a class can be mapped into multiple component interfaces.

\end{itemize}
  \vspace{-0.5em}
Figure \ref{fig:patterntocomponentmapping} shows our mapping model which maps class-to-component through FUPs. According to this mapping model, we propose the following process to mine components from APIs (c.f. Figure \ref{fig:miningprocess}):
  \vspace{-0.5em}

\begin{figure}[h]
  \begin{center}
    \includegraphics[scale=0.50]{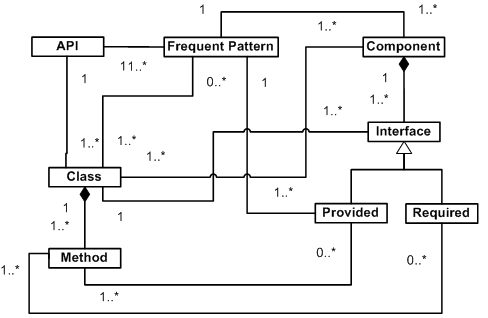}
    \caption{Mapping class to component through FUP}
        \label{fig:patterntocomponentmapping}
  \end{center}
  \vspace{-2.5em}
\end{figure}

\begin{figure}
    \vspace{-1.4em}
\centering
\parbox{5cm}{
\includegraphics[width=5cm]{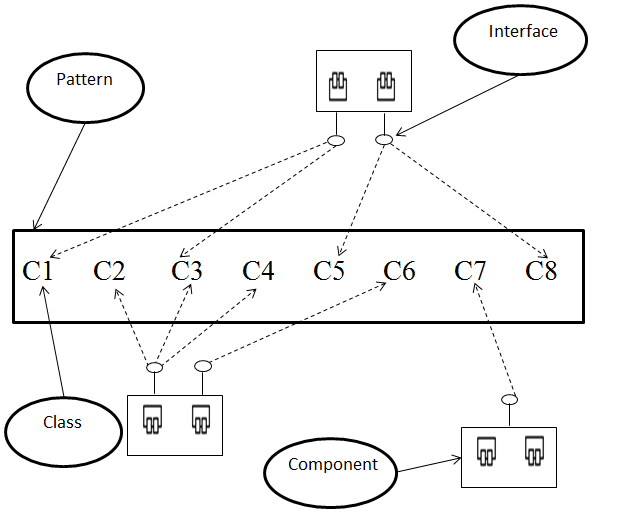}
    \caption{From FUP to provided interfaces}
\label{fig:patternToComp}}
\qquad
\begin{minipage}{5cm}
\includegraphics[width=5cm]{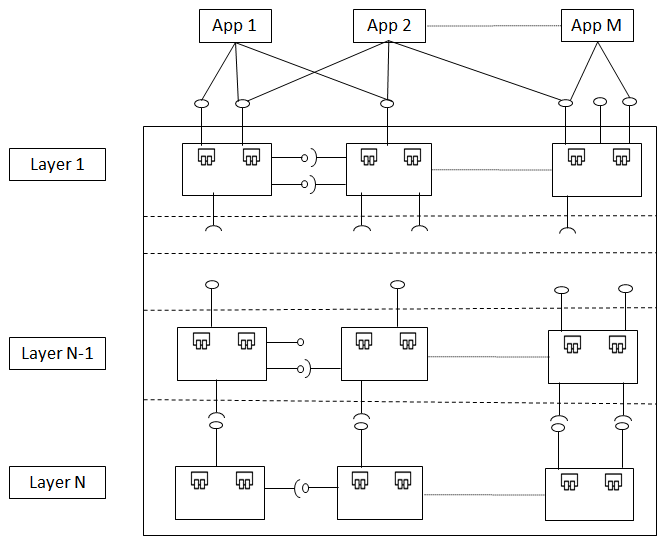}
    \caption{Multi-layers component-based API}
\label{fig:apiLayers}
\end{minipage}  
    \vspace{-2.0em}
\end{figure}

\begin{itemize}

\item \textbf{Identification of frequent usage patterns.}  FUPs are identified by analyzing the interactions between the API and its application clients.

\item \textbf{Identification of the interfaces of components.} We partition the set of classes of each FUP in subgroups, where each is considered as related to the provided interfaces of one component (c.f. Figure \ref{fig:patternToComp}). The partitioning is based on criteria related to dependencies and lexical similarity of classes and their frequency of simultaneous reuse.

\item \textbf{Identification of internal classes of components driven by their provided interfaces.} Classes forming the provided interfaces of a component form the starting point for identifying the rest of the component classes. To identify these classes we rely on the analysis of structural and behavioral dependencies between classes in the API with those forming the interfaces. We check if these classes are able to form a quality-centric component.

\item \textbf{Organizing API as Layers of Components.} 
As we previously mentioned, the API classes can be categorized according to whether they are directly reused by the API clients or not. Classes that are not directly used by API clients can also be organized into two categories. This is based on whether they belong to components identified from the classes that are directly used by API clients or not. As each class of the API must be a part of at least one component, we associate classes that do not compose any of the already identified components to new ones. Based on that, we organize component-based APIs as a set of layers describing how their components are organized. This organization is used-driven. The first layer is composed of components that are used by the software clients, while the second layer is composed of components that provide services used by components of the first layer, and so on. As a result, the API is structured in $N$ layers of components (c.f. Figure \ref{fig:apiLayers}).

\end{itemize}
  \vspace{-1.3em}

\section{Identification of component interfaces}
    \vspace{-0.7em}
The identification of classes forming an API component is driven by the identification of classes composing the provided interfaces of this component. Classes composing these interfaces are those directly accessed by the clients of the API. Classes belonging to the same interface are those frequently used together. Therefore they are identified from frequent usage patterns. Classes of the API composing frequent usage patterns are identified based on the analysis of how API classes were used by the API clients. API classes used together constitute transactions of usage.

\subsection{Extracting Transactions of Usage}
\begin{figure}
\vspace{-2.1em}
\centering
\parbox{5cm}{
\includegraphics[width=5cm]{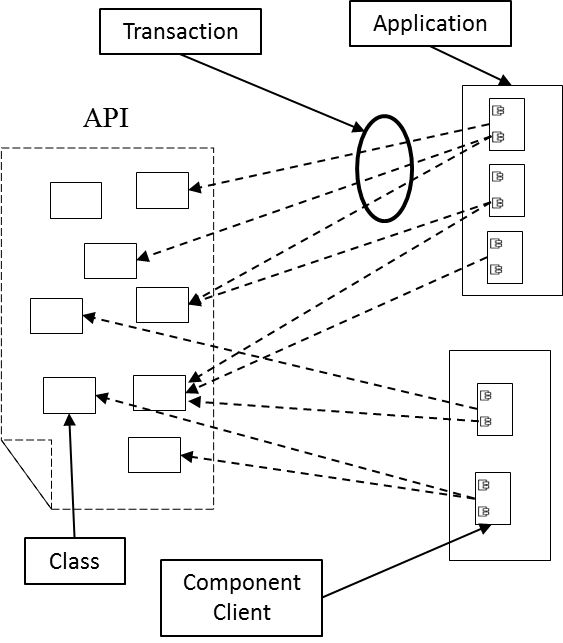}
    \caption{Client components  using API}
\label{fig:clientCom}}
\qquad
\begin{minipage}{5cm}
\includegraphics[width=5cm]{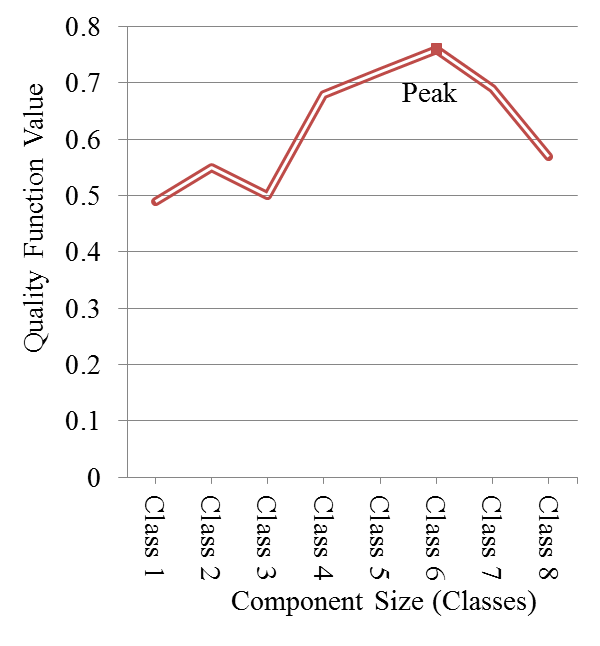}
    \caption{Identifying classes composing components}
\label{fig:potcomexample}
\end{minipage}
  \vspace{-2.5em}
\end{figure}

    \vspace{-0.8em}
A transaction of usage is a set of interactions between an API and a client of this API. These interactions consist of calling methods, accessing attributes, inheritance or creating an instance object based on a class of the API. They are identified by statically analyzing the source code of the API and its clients. Transactions are different depending on the choice of API clients. Therefore the choice of the API clients directly affects the type of the resulting patterns. A client can be considered either a class, a group of classes or the whole software application. We define a client as group of classes forming a functional component in software applications. The idea behind that is to mine patterns related to functionalities composing the applications. Thus, a transaction is a set of API classes used by classes composing a client component (c.f. Figure \ref{fig:clientCom}). To this end, we use ROMANTIC approach to identify client components composing software applications.
Algorithm \ref{algo:identifyTransactions} shows the process of transaction identification. It starts by partitioning each software client into components. Then, for each component, it identifies API classes that are reused by the component classes.
  \vspace{-1.3em}
\subsection{Mining Frequent Usage Patterns of Classes}
  \vspace{-0.7em}
In the previous step, the interactions of all client components with the API are identified as transactions. Based on these transactions, we identify FUPs. A FUP is a set of API classes that are frequently used together by client components. It allows the detection of hidden correlations of usage among classes of the API. We mine FUPs based on the FPGrowth algorithm \cite{han2006data}. In this algorithm, a pattern is considered as frequent if it reaches a predefined threshold of interestingness metric. This metric is known as $Support$. The $Support$ refers to the probability of finding a set of API classes in the transactions. The use of the $Support$ metric separates the classes of API into two groups according to whether they belong to at least one FUP or not. Classes that do not belong to any of the identified FUPs are the less commonly used classes. As each API class that belongs to a transaction is a class that has been accessed by the clients of the API, therefore it must be a part of the classes composing the interfaces of at least one component. We propose assigning each class of the less commonly used classes to the pattern holding the maximum $Support$ value when they are merged together.

\vspace{-0.8em}
\subsection{Identifying Classes Composing Component Interfaces from Frequent Usage Patterns}
  \vspace{-0.7em}
We identify classes composing component interfaces from those composing FUPs. Each FUP is partitioned into a set of groups, where each group represents a component interface. Classes are grouped together according to three heuristics that measure the probability of a set of classes to be a part of the same interface. The first is the frequency of simultaneous use of these classes by the same client. The second is the cohesion of these classes. This measures the strength of sharing data (e.g. attributes) between these classes. The third heuristic is the lexical similarity between these classes based on the textual names of the classes, their methods as well as their attributes.
Based on the above heuristics, we propose a fitness function, given below, measuring the ability of a group of classes to form a component interface. We use \textit{LCC} metric \cite{Bieman:1995:CRO} to measure the cohesion of a set of classes, \textit{Conceptual Coupling} metric \cite{Poshyvanyk:2006:ICSM} to measure classes' lexical similarity and  \textit{Support} metric to measure the association frequency of a set of classes.
The partition of each pattern into groups of classes is based on a hierarchical clustering algorithm which uses this function as a function of similarity.

\begin{equation}
I(E)=\frac{1}{\sum_{i}\lambda_i}\cdot(\lambda_1\cdot LCC(E) + \lambda_2\cdot CC(E) + \lambda_3\cdot S(E))
\end{equation}

Where \textit{E} is a set of OO classes; \textit{LCC(E)} is the \textit{Cohesion} of \textit{E}; \textit{CC(E)} is \textit{Conceptual Coupling} of \textit{E}; \textit{S(E)} is the \textit{Support} of \textit{E}; and $\lambda_1$, $\lambda_2$, $\lambda_3$ are weight values, situated in [0-1]. These are used by the API expert to weight each characteristic as needed.
\vspace{-1.5em}
\begin{algorithm}

\SetAlgoLined
\caption{Identifying Transactions}
\label{algo:identifyTransactions}
\SetKwFunction{match}{match}
 \KwIn{Source Code of a Set of Software Clients($clients$), API Source Code($api$)}
 \KwOut{A Set of Transactions($trans$)}
 \For{each $c$ in $clients$}
 {
  $components$.add(ROMANTIC($c$));
 }
 \For{each $com$ in $components$}
 {
 $transaction$ = $\varnothing$;\\
   \For{each $class$ in $com$}
    {
     $transaction$.add(class.getUsedClasses($api$));
    }
    $trans$.add($transaction$);
   }
\KwRet{$trans$}\\
\end{algorithm}

  \vspace{-3.0em}
\section{API as Library of Components }
\vspace{-0.5em}
\subsection{Identifying Classes Composing Components}
\label{firstlevelcomponent}
\vspace{-0.5em}
As we mentioned before, the component identification process is driven by the identification of its provided interfaces. This means that API classes forming a component are identified in relation to their direct or indirect structural and behavioral dependencies with the classes forming provided interfaces of the component.
The selection of a class of the API to be a part of the component classes is based on the measurement of the quality of this component, when this is included.
The identification of these classes is done gradually. In other words, we start to form the group of classes composing the interface ones, and then we add other classes to form a component based on the component quality measurement model. Classes having either direct or indirect links with the interface ones represent the candidate classes to be added to them. At each step, we add a new API class. This is selected based on the quality value of the component, formed by adding this class to the ones already selected. The class that maximizes the quality value is selected in this step. This is done until all API classes are investigated. Each time we add a class, we evaluate the component quality. Then, we select the peak quality value to decide which classes form the component. This means that we exclude classes added after the peak value. As an example, $Class7$ and $Class8$ in Figure \ref{fig:potcomexample} are excluded from the resulting component because they were added after the quality value reached the peak. 

\vspace{-1.0em}
\subsection{Organizing API as Layers of  Components}
\vspace{-0.5em}
As we previously mentioned, the API is structured in $N$ layers of components.
To identify components of layer $L$, we rely on components of layer $L-1$. We proceed similarly to the identification of the components of the first layer. We use required interfaces of the components already identified in layer $L-1$ to identify the interfaces provided by components in layer $L$. This continues until reaching a layer where its components either do not require any interface or they require ones already identified.
Each interface defined as a required for a component of layer $L-1$ is considered as a provided by a component of layer $L$ except ones provided by the already identified components. All interfaces provided in layer $L$ are grouped into clusters to identify those provided by the same component of layer $L$. The clusters are obtained based on a hierarchical clustering algorithm. This algorithm uses a similarity function that measures: (i) the cohesion of classes composing a group of interfaces, (ii) the lexical similarity of these classes and (iii) the frequency of their simultaneous use. Clusters that maximize this function are selected. The interfaces composing each cluster are considered as provided by the same component. Analogously to the identification of the components of the first layer, the other classes composing each component are identified starting from classes composed of its already identified provided interfaces. 
\vspace{-1.3em}
\section{Experimentation and Results}
\vspace{-0.5em}
\subsection{Experimental Design}
\vspace{-0.5em}
\subsubsection{Data Collection}
We collected a set of 100 $Android-Java$ applications from open-source repositories\footnote{$sourceforge.net$, $code.google.com$, $github.com$, $gitorious.org$, and $aopensource.com$}. The average size of these applications in terms of number of classes is 90. The application names are listed in the Appendix. These applications are developed based on classes of the $android$ APIs\footnote{We select android API level 14 as a reference}. In our experimentation, we focus on three of these APIs.  The first is the $android.view$ composed of 491 classes. This API provides services related to the definition  and management of the user interfaces in android applications. The second API is the $android.app$ composed of 361 classes. This API provides services related to creating and managing android applications. The last API is the $android$ composed of 5790 classes. This API includes all of the android services.
\vspace{-1.0em}
\subsubsection{Research Questions and Evaluation Method}
The approach is evaluated on the collected software applications and APIs. We identify client components independently from each software application. Each component in software is considered as a client of the APIs to form a transaction of classes. Then, we mine Frequent Usage Pattern (FUP) from the identified transactions. Next, from classes composing each FUP, we identify classes composing a set of component interfaces. Then, we identify all component classes starting from ones composing their interfaces. Lastly, the final results obtained by our approach are presented.

We evaluate the obtained components by answering the three following research questions.
\begin{itemize}
\item \textbf{RQ1: Does the Approach Reduce the Understandability Efforts?}
This research question aims at measuring the saved efforts in the API understandability that are brought by migrating object-oriented APIs into component-based ones.

\item \textbf{RQ2: Are the Mined Components Reusable?}
As our approach aims at mining reusable components, we evaluate the reusability of the resulted component. This is based on measuring how much related classes are grouped into the same components.

\item \textbf{RQ3: Is the Identification of Provided Interfaces Based on FUPs Useful?}
The proposed approach identifies the provided interfaces of the components based on how clients have used the API classes (i.e. FUPs). Thus, this research question evaluates how much benefit the use of FUPs brings by comparing components identified by our approach with ones identified without taking FUPs into account.
\end{itemize}
\vspace{-0.5em}
To answer the second question that related to the reusability, we use the $K-fold$ cross validation method \cite{han2006data}. The idea is to partition the client applications into $K$ parts. Then, the  identification process is applied $K$ times by considering, each time, $K-1$ different parts for the identification process and by using the remaining part to measure the reusability. Next, we take the average of all $K$ trail results. In our experiment, we set $K$ to 2, 4, and 8.

\begin{figure}[h]
\vspace{-2.2em}
  \begin{center}
    \includegraphics[scale=0.42]{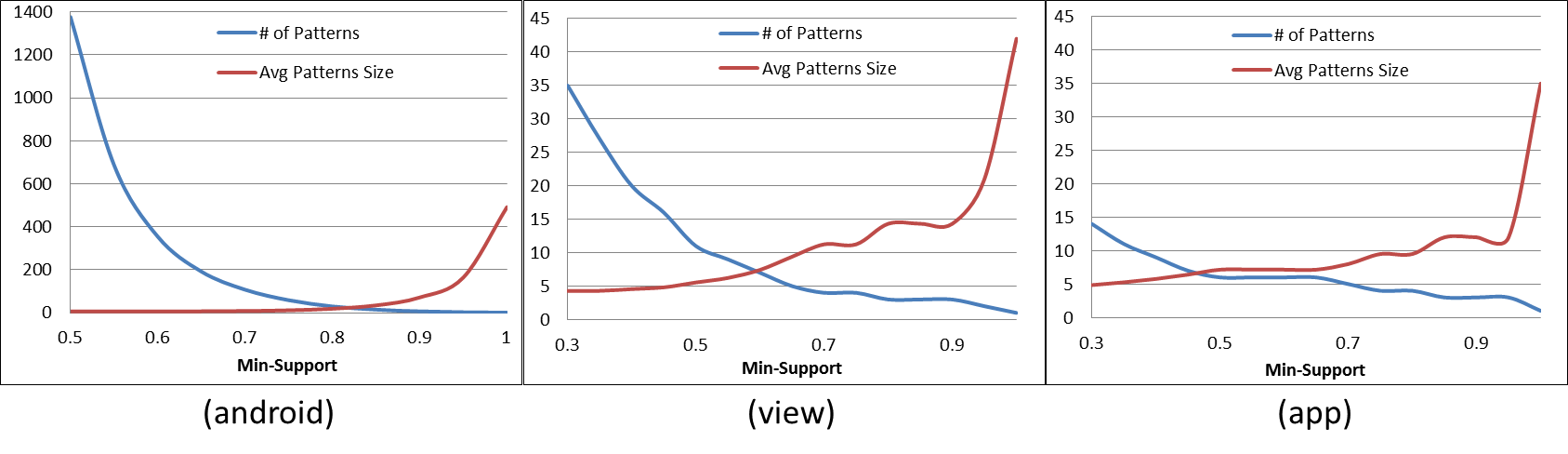}
    \caption{Changing the support threshold value to mine FUPs of classes}
        \label{fig:changeSupportResults}
  \end{center}
  \vspace{-4.0em}
\end{figure}

\subsection{Results}
\vspace{-0.4em}
\subsubsection{Intermediate Results and Identified Components}
The average number of client components identified from each software is $4.5$ and the average number of classes composing each component is $18.73$. Table \ref{transactionsResults} shows the average number of transactions per software application ($ANTIC$), the average of transaction size in terms of classes ($ATS$), and the percentage of components that have used the API ($PCU$). The last column of this table shows an example of transactions.

The results show that $android$, $view$, and $app$ APIs have been used respectively by only 54\%, 29\% and 32\% of client components. In addition, we note that each client component has used the API classes intensively compared to the number of classes composing it. For example, the transaction size is 17.91 classes for the $view$ API, where the average number of classes per component is 18.73. This is due to the fact that classes that serve the same services in software applications, and consequently depend on the same API classes, are grouped together in the same client component.

\begin{table}[ht]
\vspace{-2.0em}
\centering
\caption{The Identification of Transactions  }
\begin{tabular}{|c|c|c|c|p{8.2 cm}|} \hline
API & ANTIC & ATS & PCU  & Example\\\hline
$android$&	2.61&	64.82 & 0.54 & Bitmap, Path, Log, Activity, Location, Canvas, Paint, ViewGroup, MotionEvent, View, TextView, GestureDetector\\ \hline
$view$&	1.51&	17.91& 0.29 & MenuItem, Menu, View, ContextMenu, WindowManager, MenuInflater, Display, LayoutInflater  \\ \hline
$app$&	1.58&	10.90& 0.32& ProgressDialog, Dialog, AlertDialog, Activity, ActionBar, Builder, ListActivity\\ \hline
\end{tabular}
\label{transactionsResults}
\vspace{-2.0em}
\end{table}

The identification of  FUPs relies on the value of the $Support$ threshold. The number and the size of the mined FUPs depend on this value. For all application domains where FUPs are used (e.g. data mining), this value is determined by domain experts.
In our approach, to help API experts to determine this value, we assign the $Support$ threshold values situated in [30\%-100\%]. We give for each $Support$ value the number of the mined FUPs and the average size of the mined FUPs for each API (c.f. Figure \ref{fig:changeSupportResults}). The results show that the number of mined FUPs is directly proportional to the $Support$ value, while the average size of the mined FUPs is inversely proportional.

Based on their knowledge of the API, API experts can select the value of the $Support$. For example, if the known average number of API classes used together to implement an application service is $N$, then the experts can choose the $Support$ value corresponding to FUPs having $N$ as the average size.
Based on the obtained results and our knowledge on android APIs\footnote{The authors of this paper are experts on the android APIs}, we select the $Support$ threshold values as 60\%, 45\%, and 45\% respectively for the $android$, the $view$ and the $app$ APIs.

Table \ref{patternExamples} shows examples of the mined FUPs. For instance, the FUP related to $view$ API contains 10 classes. The analysis of this FUP shows that it corresponds to three services: animation (\textit{Animation} and \textit{AnimationUtils} classes), view (\textit{Surface, SurfaceView, SurfaceHolder, MeasureSpec, ViewManager} and \textit{MenuInflater} classes), and persistence of the view states (\textit{AbsSavedState} and \textit{AccessibilityRecord} classes). These services are dependent. Animation service needs the view service and the data of animation view needs to be persistent.

\begin{table}[ht]
\vspace{-2.0em}
\centering
\caption{Examples of the Mined FUPs}
\begin{tabular}{|c|p{10.8 cm}|} \hline
API & Example\\\hline
$android$& Intent, Context, Log, SharedPreferences, View, TextView, Toast, Activity, Resources
\\ \hline
$view$&	Surface, Animation, AnimationUtils, AccessibilityRecord, ViewManager, MenuInflater, AbsSavedState, SurfaceView, SurfaceHolder, MeasureSpec\\ \hline
$app$& Dialog, Activity, ProgressDialog\\ \hline
\end{tabular}
\label{patternExamples}
\vspace{-2.0em}
\end{table}

In Table \ref{patternToInterfaceResults}, we present the results of interface identification in terms of the average number of component interfaces identified from a FUP ($ANCIP$), the average number of classes composing component interfaces ($ACIS$) and the total number of component interfaces in the API ($TNCI$). The last column of this table presents examples of component interfaces identified from the FUPs given in Table \ref{patternExamples}. For instance, the analysis of classes composing the component interfaces identified from the FUP related to the $view$ API shows that they are related to surface view services.

\begin{table}[ht]
\vspace{-2.0em}
\centering
\caption{Identification of Component Interfaces from FUPs}
\begin{tabular}{|c|c|c|c|c|} \hline
API & ANCIP & ACIS & TNCI & Examples\\ \hline
$android$&	1.57&	5.62&	232 & Activity, View, TextView, Toast\\ \hline
$view$&	2.17&	2.94&	19 & Surface, SurfaceView, SurfaceHolder\\ \hline
$app$&2.50&	4&	10& Dialog, ProgressDialog\\ \hline
\end{tabular}
\label{patternToInterfaceResults}
\vspace{-2.0em}
\end{table}

Table \ref{potentialComResults} presents the results related to the mined components composing the first API layer. For each API, we give the number of the mined components ($NMC$) and the average number of classes composing the mined components ($ACS$). The last column of this table shows examples of classes composing components identified started from classes composing provided component interfaces presented in Table \ref{patternToInterfaceResults}. The results show that the services offered by classes of $android$, $view$ and $app$ APIs are identified as 232, 19 and 10 components respectively. This means that developers only require to interact with these components to get the needed services from these APIs.

\begin{table}[ht]
\vspace{-1.0em}
\centering
\caption{Identifying Classes Composing Components}
\begin{tabular}{|c|c|c|p{9.3 cm}|} \hline
API & NMC & ACS & Example \\ \hline
$android$&	232&	19.99 & Activity, View, TextView, Toast, Drawable, GroupView, Window, Context, ColorStateList, LayoutInflater\\ \hline
$view$&	19&	7.49 & Surface,SurfaceView, SurfaceHolder, MockView, Display, CallBack \\ \hline
$app$&	10&	5.86 & Dialog, ProgressDialog, AlertDialog\\ \hline
\end{tabular}
\label{potentialComResults}
\vspace{-2.0em}
\end{table}

Table \ref{FinalResults} shows the final results obtained by our approach. For each API, we firstly give the size of the API in terms of the number of OO classes composing the API and the number of the identified components. Secondly, we present the total number of used entities (classes and respectively components) by the software clients. The results show that classes participating in providing related services are grouped into one component. Furthermore, the total number of cohesive and decoupled services is identified for each API. For instance, $android$ API consists of 497 components (coarse-grained services), while $view$ and $app$ APIs contain 43 and 55 components respectively.

 \begin{table}[ht]
 \vspace{-2.0em}
     \centering
     \caption{The Final Results}
     \begin{tabular}{|c|c|c|c|} \hline
     API Name & API Entity & API size & No. of used Entities\\\hline
     \multirow{2}{*} {$android$}& $OO$&	5790&	491\\ \cline{2-4}
     & $CB$&	497&	54\\ \hline
     \multirow{2}{*} {$view$} & $OO$&	491&	42\\ \cline{2-4}
     & $CB$&	43&	17\\ \hline
      \multirow{2}{*}{$app$}& $OO$&	361&	45\\ \cline{2-4}
     & $CB$&	55&	5\\ \hline
     \end{tabular}
     \label{FinalResults}
     \vspace{-3.5em}
     \end{table}

\subsubsection{Answering Research Questions}
\vspace{0.70mm}
\textit{RQ1: Does the Approach Reduce the Understandability Efforts?}
The efforts spent to understand such an API is directly proportional to the complexity of the API. This complexity is related to the number of API elements and the individual element's complexity. On the one hand, the reduction in the number of elements composing the API is obtained by grouping classes collaborating to provide one coarse-grained service into one component. The results show that the average number of identified components for the studied APIs is 11\% ( ( (497/5790) + (43/491) + (55/361) ) /3 ) of the number of classes composing the APIs. This means that the API size is significantly reduced by mapping class-to-component. On the other hand, the reduction in the individual element complexity is done by migrating object-oriented APIs into component-based ones. Meaning, components define their required and provided interfaces, while OO classes at least do not define required interfaces (e.g. a class may call a large number of methods belonging to a set of classes without an explicit specification of these dependencies). The results show that the average number of used components for the APIs is 4\% ( ( (54/491) + (17/42) + (5/45) ) /3 ) of the number of used classes. This means that the effort spent to understand API entities is significantly reduced in the case of software applications developed based on API components compared to the development based on API classes. Note that, developers only need to understand the component interfaces, but not the whole component implementation.

\textit{RQ2: Are the Mined Components Reusable?}
We consider that the reusability of a software component is related to the number of used classes among all ones composing the software component. Thus, we calculate the reusability of the component based on the ratio between the numbers of used classes composing the component to the total number of classes composing the component. To prove that our resulted component-based APIs could be generalized to another independent set of client applications, we rely on $K-fold$ cross validation method. Table \ref{ValidationResults} presents the results of this measurement. These results show that the reusability results is distributed in a disparate manner. The reason behind this disputation is the size of the train and test data as well as the size of the API. For instance, the average reusability for the $app$ API is 37\% when the number of train clients is 50 application clients, while it is 51\% when the number of train clients is 88 application clients. Thus, the reusability of the components increases when the number of train client applications increases. The results show that our approach identifies reusable components, where the average reusability for all APIs is 47\%.

\begin{table}
 \vspace{-2.0em}
          \centering
          \caption{Reusability Results }
  \begin{tabular}{|c|c|c|c|c|c|c|c|c|c|}
    \hline
    API &
      \multicolumn{3}{|c|}{$android$} &
      \multicolumn{3}{|c|}{$view$} &
      \multicolumn{3}{|c|}{$app$} \\
    \hline
    K & 2 & 4 & 8 & 2 & 4 & 8 & 2 & 4 & 8 \\
    \hline
    Reusability & 40\% & 43\% & 57\% & 46\% & 48\% & 56\% & 37\% & 41\% & 51\%\\
    \hline
  \end{tabular}
\label{ValidationResults}
\vspace{-2.0em}
\end{table}

\textit{RQ3: Is the Identification of Provided Interfaces Based on FUPs Useful?}
To prove the utility of using FUPs during the identification process, we compare the components mined based on our approach with ones mined using ROMANTIC approach, which does not take FUPs into consideration. This is based on the density of using the component provided interfaces by application clients. The density refers to the ratio between the number of used interface classes to the total number of interface classes for each component. Table \ref{ProvidedInterfacesValidation} shows the average density for the two identification approaches. These results show that our approach outperforms ROMANTIC approach. For instance, the application clients need to reuse a larger number of components of ones mined based on ROMANTIC with less density of provided interface classes compared to component mined based on our approach. For instance, the average usage density of classes composing provided interfaces of ROMANTIC components is 21\%, while it is 61\% for components mined by our approach for all APIs.

\begin{table}[ht]
 \vspace{-2.0em}
\centering
\caption{The Results of Interface Density}
\begin{tabular}{|c|c|c|} \hline
API & ROMANTIC & Our Approach \\ \hline
$android$& 20\%	& 69\%\\ \hline
$view$&	18\%&	58\%\\ \hline
$app$& 26\%& 55\%\\ \hline
\end{tabular}
\label{ProvidedInterfacesValidation}
 \vspace{-3.0em}
\end{table}

\section{Related work}
\vspace{-0.6em}
\label{relatedWork}
To the best of our knowledge, no approach has been proposed to identify components from object-oriented APIs. However, we present two research areas that are related to our approach. The First one aims at identifying components from OO software applications. The second area aims at mining frequent patterns of API usage. 

Concerning the identification of software components from the source code of software applications, numerous approaches have been presented. 
Garcia et al. provide a survey of some of these approaches  \cite{Garcia:ASE:2013}. In \cite{Detten:2013:reengineering}, Detten et al. presented the Archimetrix approach, which aims at mining the architecture of the legacy software. It relies on a clustering algorithm to partition the system classes into components. This algorithm depends on name resemblance, coupling and cohesion metrics as a fitness function. In \cite{Kebir:2012:WICSA}, Kebir et al. presented an approach to extract components from a single OO software system. Classes composing the extracted components form a partition. Mined components are considered as a part of the component-based architecture of the corresponding software. In \cite{allier:2011:WICSA} Allier et al. depended on dynamic dependencies between classes to recover components. Based on the use case diagram, the execution trace scenarios are identified. Classes that frequently occur in the execution traces are grouped into a single component. Cohesion and coupling metrics are also taken into account during the identification process. Weinreich et al. proposed, in \cite{Weinreich:2012:WICSA}, an approach to recover multi-view architecture model of software applications implemented based on service oriented architecture. The authors classified software artifacts based on the information from source code, configuration files and binary codes. 
In \cite{shatnawi:2013:IRI}, an approach has been presented to mine reusable components from a set of similar software applications. A component is considered as more reusable, when it is reused many times by the software applications. The authors firstly identified components independently from each software application. Then, based on the lexical similarity between the classes composing these components, they identified reusable ones. 

In the context of API mining, many approaches have been proposed to mine frequent usage patterns of APIs based on the usage history of APIs. Robillard et al. provide a survey of these approaches  \cite{Robillard:2013:IEEETran}. These approaches can be mainly classified based on four main criteria. The first one is related to the goal, which can be either giving examples and recommendations of how to use API entities (e.g. \cite{Montandon:2013:WCRE,Uddin:2012:TAA}), supporting the documentation of APIs (e.g. \cite{Montandon:2013:WCRE,Wang:2013:MSH}), or improving the bug detection task (e.g. \cite{Monperrus:2010:ECOOP}).
The second criterion is related to the pattern ordering, where some approaches mine ordered patterns (e.g.  \cite{Montandon:2013:WCRE,Wang:2013:MSH}), while other ones mine unordered patterns (e.g. \cite{Monperrus:2010:ECOOP,Bruch:2006:FIS}). The third one concerns the granularity of the elements composing a pattern. For examples, in \cite{Montandon:2013:WCRE,Wang:2013:MSH}), the approaches mine patterns composed of methods, and the approach in \cite{Bruch:2006:FIS} mines patterns composed of classes. The fourth one related to the technique that is used to identify the patterns. The used technique can be association rules mining (e.g. \cite{Bruch:2006:FIS}), clustering algorithms (e.g. \cite{Wang:2013:MSH}) or a heuristic defined by the authors such as \cite{Montandon:2013:WCRE,Monperrus:2010:ECOOP}. Some approaches combines many techniques, e.g., Unddin et al. used Principle Component Analysis with Clustering algorithm \cite{Uddin:2012:TAA}, and Buse and Weimer combined the clustering algorithm with their own proposed heuristic \cite{Buse:2012:SAU}.
 \vspace{-1.0em}
\section{Conclusion and Future Work}
 \vspace{-0.5em}
In this paper, we presented an approach aimed at mining software components from object-oriented APIs. It is based on static analysis of the source code of both the APIs and their software clients. The mining process is used-driven. This means that components are identified starting from classes composing their interfaces. Classes composing the provided interface of the first layer components compose FUPs. We experimented our approach by applying it on a set of open source $Java$ applications as clients for three android APIs. The results show that our approach improves the reusability of the API.

As our approach is used-driven, the results depend on the quality and the number of usages of the API. This means that identified FUPs rely on the considered software clients. Therefore the identification of provided interfaces and then their corresponding components depends on API clients. Consequently it is essential to select clients having the largest number of usages of the API.

Our future work will focus on migrating the identified OO components into existing component models such as OSGI model, and developing a visual environment that allows domain experts to interact with the approach at each step of the identification process, thus modify the obtained results as needed.

 \vspace{-1.0em}
\section*{Appendix:}
 \vspace{-0.5em}
These are the names of the applications that considered as clients of the APIs.

\textit{ADW Launcher, APV, ARMarker, ARviewer, Alerts,
Alogcat, AndorsTrail, AndroMaze, AndroidomaticKeyer, AppsOrganizer,
AripucaTracker, AsciiCam, Asqare, AugmentRealityFW, AussieWeatherRadar,
AutoAnswer, Avare, BansheeRemote, BiSMoClient, BigPlanetTracks,
BinauralBeats, Blokish, BostonBusMap, CalendarPicker, CH-EtherDroid, CVox, CamTimer, ChanImageBrowser, CidrCalculator, ColorPicker,
CompareMyDinner, ConnectBot, CorporateAddressBook, Countdown, CountdownTimer,
CrossWord, CustomMaps, DIYgenomics, Dazzle, Dialer2,
DiskUsage, DistLibrary, Dolphin, Doom, DriSMo,
DroidLife, DroidStack, Droidar, ExchangeOWA, FeedGoal,
FileManager, FloatingImage, Gcstar, GeekList, GetARobotVPNFrontend,
GlTron, GoHome, GoogleMapsSupport, GraphView, HeartSong,
Hermit, Historify, Holoken, HotDeath, Introspy,
LegoMindstroms, Lexic, LibVoyager, LiveMusic, LocaleBridge, MAME4droid, Look, LookSocial, Macnos, Mandelbrot,
Mathdoku, MediaPlayer, Ministocks, MotionDetection, NGNStack,
NewspaperPuzzles, OnMyWay, OpenIntents, OpenMap, OpenSudoku,
Pedometer, Phoenix, PhotSpot, Prey, PubkeyGenerator,
PwdHash, QueueMan, RateBeerMobile, AlienbloodBath, SuperGenPass,
SwallowCatcher, Swiftp, Tumblife, VectorPinball, WordSearch.}

%

%
%


\end{document}